\documentclass[notitlepage,floatfix,onecolumn,aps,showpacs,superscriptaddress,longbibliography,10]{revtex4-1}

\usepackage{graphicx}
\usepackage{multirow,booktabs}
\usepackage{amsmath,amssymb,latexsym,MnSymbol}
\usepackage[colorlinks=true, urlcolor=black,linkcolor=black,citecolor=blue]{hyperref}

\makeatletter
\let\cat@comma@active\@empty
\makeatother

\usepackage{lineno}
\usepackage{verbatim}
\begin{document}

\title{Unfolding the Complexity of the Global Value Chain: Strength and Entropy in the Single-Layer, Multiplex, and Multi-Layer International Trade Networks}

\author{Luiz G. A. Alves}
\affiliation{Institute of Mathematics and Computer Science, University of S\~ao Paulo, S\~ao Carlos, SP 13566-590, Brazil}

\author{Giuseppe Mangioni}
\affiliation{Dipartimento di Ingegneria Elettrica, Elettronica e Informatica, University of Catania, 95125 Catania, Italy}

\author{ Francisco A. Rodrigues}
\affiliation{Institute of Mathematics and Computer Science, University of S\~ao Paulo, S\~ao Carlos, SP 13566-590, Brazil}
\affiliation{Mathematics Institute, University of Warwick, Gibbet Hill Road, Coventry CV4 7AL, UK}
\affiliation{Centre for Complexity Science, University of Warwick, Coventry CV4 7AL, UK}

\author{Pietro Panzarasa}
\affiliation{School of Business and Management, Queen Mary University of London, London E1 4NS, UK}

\author{Yamir Moreno}\email{ yamir.moreno@gmail.com}
\affiliation{Department of Theoretical Physics, University of Zaragoza, 50009 Zaragoza, Spain}
\affiliation{Institute for Biocomputation and Physics of Complex Systems, University of Zaragoza, \mbox{50009 Zaragoza, Spain} }
\affiliation{The Institute for Scientific Interchange Foundation, 10126 Torino, Italy}

\begin{abstract}
The worldwide trade network has been widely studied through different data sets and network representations with a view to better understanding interactions among countries and products. Here we investigate international trade through the lenses of the single-layer, multiplex, and multi-layer networks. We discuss differences among the three network frameworks in terms of their relative advantages in capturing salient topological features of trade. We draw on the World Input-Output Database to build the three networks. We then uncover sources of heterogeneity in the way strength is allocated among countries and transactions by computing the strength distribution and entropy in each network. Additionally, we trace how entropy evolved, and show how the observed peaks can be associated with the onset of the global economic downturn. Findings suggest how more complex representations of trade, such as the multi-layer network, enable us to disambiguate the distinct roles of intra- and cross-industry transactions in driving the evolution of entropy at a more aggregate level. We discuss our results and the implications of our comparative analysis of networks for research on international trade and other empirical domains across the natural and social sciences. 
\end{abstract}

\maketitle

\section{Introduction}


Network perspectives have long been advocated to investigate international trade among countries~\cite{allard2017geometric,caldarelli2012network,caldarelli2018physics,cingolani2017countries,cristelli2015heterogeneous,
de2011world,fagiolo2009world,formichini2018measuring,garlaschelli2004fitness,garlaschelli2005structure,he2010structure,
hidalgo2007product,hidalgo2009building,serrano2003topology,tacchella2012new}, and to study several other economic systems, including the global economy of transnational credit and
investments, supply chains and production markets~\cite{schweitzer2009economic}. The idea of representing the worldwide trade system as a complex network can be traced back to the 20th century~\cite{de2011world}, when it was used to describe the patterns of international transactions among countries before World War II. Different frameworks have helped researchers to better understand how countries interact with each other in the different industries of the economy as well as to uncover the intricacies that might affect countries' competitive advantage and economic performance~\cite{hidalgo2009building,tacchella2012new}. 

Different network models and representations have been proposed to describe the structure of the worldwide trade network. The most common, perhaps, is the single-layer framework~\cite{de2011world}, where the countries are represented as the nodes of the network and economic transactions as directed edges that are established from (to) one country to (from) another when the former exports (imports) products of services to (from) the latter. Using this approach, researchers have shed light on a number of structural features of trade. For example, it has been suggested that the worldwide trade network is characterized by a community structure~\cite{barigozzi2011community,barigozzi2011identifying,piccardi2012existence}, a heavy-tailed degree distribution~\cite{fagiolo2009world}, and small-world and clustering properties~\cite{serrano2003topology}. Another widely adopted representation of trade is the bipartite network, which is a special case of the single-layer network characterized by two sets of distinct nodes (e.g., countries and products) such that connections are only allowed between nodes in different sets (e.g., a country to the products exported or imported)~\cite{newman2018networks}. For instance, using the bipartite representation, it has been shown that topological changes of the network can be regarded as early signals of economic downturns in the worldwide trade~\cite{saracco2015randomizing}. Additionally, researchers have used the bipartite network representation to unveil the competitiveness of countries and the complexity of products from trade data~\cite{cristelli2013measuring,hidalgo2009building}. 

Recently, there has been a growing interest in examining several economic systems using multiplex and multi-layer networks~\cite{battiston2014structural,kivela2014multilayer}. Because international trade unfolds within and across industries and the countries can be involved in multiple stages of production along the global value chain, using network frameworks in which industries can be represented as the layers of the network and connections among countries can be investigated through the lenses of such layers can offer invaluable insights into the structure of trade. For instance, recent studies have adopted a multiplex network perspective to investigate the role of layer-specific local constraints in shaping international trade~\cite{mastrandrea2014reconstructing}, to simulate the emergence and unfolding of cascading failures in the trade system~\cite{lee2016strength}, to study the structural position and influential role of countries based on betweenness centrality measures~\cite{ghariblou2017shortest}. More recently, other studies have drawn on multi-layer networks to investigate the structural organization of the worldwide trade~\cite{alves2018nested} and the impact of technological innovations on industrial products~\cite{formichini2018measuring}.

A network representation should be adopted when it has clear advantages over the others regarding the quality of the information that can be extracted from trade data and the suitability of the framework for capturing salient topological features of the underlying structure and dynamics of trade. 
While multiplex and multi-layer networks have been increasingly used to study trade over recent years, the actual benefits of using more complex network structures over simpler and more parsimonious ones still remain to be fully investigated. In this work, we take a step in this direction and propose a general framework for investigating the relative advantages of using the single-layer, multiplex, and multi-layer networks for representing international trade. To this end, we draw on trade data to construct the global value chain of the worldwide production network. We focus on one specific network property---node strength---and show how the distribution of such property varies from one network representation to the other. We further shed light on the heterogeneity of strength in the system by computing entropy and its evolution over time, and show that different ways of aggregating economic transactions (i.e., by country, by country and industry, and by country and combinations of industries) enable different market structures and sources of heterogeneity to be~detected. 

The article is organized as follows. First, we present the data and the methods (Section~\ref{subsection:data}). We~then outline the approach to building each network representation (Section~\ref{subsection:net}). Next, in Section~\ref{section:results}, we describe a number of measures for computing the strength of nodes in each network representation, and use these measures to construct and compare the strength distributions. We further investigate differences between the three networks by measuring the entropy of strength and assessing how heterogeneously trade value is distributed in each network. Finally, we focus on the evolution of entropy in each network, and uncover the components of trade that are responsible for the overall trend observed at the aggregate level. In Section~\ref{section:dissussion}, we summarize and discuss our findings. 
 
\section{Materials and Methods}\label{section:methods}\vspace{-6pt}

\subsection{The Data Set}\label{subsection:data}

Our analysis draws on the World Input-Output Database (WIOD)~\cite{timmer2015illustrated}. This database contains information about the transactions among $43$ countries, including $28$ European Union countries and $15$ other major countries within the period from 2000 to 2014. Moreover, the database provides details on transactions within and between $56$ industries and the final demand, including transactions involving the same country within and across industries, thus enabling us to analyze the production stages along the whole global value chain. 

For every year, transactions among countries are expressed in millions of US dollars, in current prices, and a matrix representing the world input-output table (WIOT) is provided. The final demand is articulated into five distinct components: the final consumption expenditure, the~final consumption expenditure by non-profit institutions serving households (NPISH), the final consumption expenditure by government, the gross fixed capital formation, and the changes in inventories and valuable. Because our study is directly concerned with the values (weights) of transactions, and because changes in inventories can be both positive and negative over the years, we filtered out from the analysis the individual transactions involving the final demand that were associated with a total negative value. Notice that such negative values could only be found in the multi-layer network where transactions are disaggregated within and between industries. As a result of this filtering, only approximately 0.03\% of observations per year were removed. A detailed description of the data set can be found in~References~\cite{alves2018nested,timmer2015illustrated}.

\subsection{Building the Single-Layer, Multiplex, and Multi-Layer Networks}\label{subsection:net}

We started our analysis by representing the WIOT as a network. Traditionally, the worldwide trade network has been investigated as a single-layer (or monoplex) network~\cite{barigozzi2011community,barigozzi2011identifying,piccardi2012existence,saracco2015randomizing,serrano2003topology}. Thus, a first approach to studying the WIOD would be to extract information about the dynamics of the interactions among countries by focusing on the single-layer representation of the network. In this representation, a directed link is established from one country $c_i$ to another country $c_j$ if there is a transaction from $c_i$ to $c_j$. Moreover, the intensity of each economic transaction is given by the weight $w_{c_ic_j}^{S}$ of the link connecting the countries, which is equal to the total amount of value (in US dollars) exchanged in the transaction. 

Formally, we define our single-layer network as a directed graph given by $G^{S}=(V^{S},E^{S},W^{S})$, where $V^{S}=\{c_i;\,i\,\in \{1,\dots, N\}\}$ is a set of nodes, with $N$ equal to the number of countries ($N=43$), $E^{S}=\{(c_i,c_j); \,i,j\, \in \{1,\dots, N\}\}$ is a set of directed edges between pairs of nodes (including self-loops), and $W^S=\{w^{S}_{c_ic_j};\,i,j\, \in \{1,\dots, N\}\}$ is the set of the weights associated with the edges. Thus, a given element $a^{S}_{c_ic_j}$ of the adjacency matrix $A^{S}$ of the single-layer network can be defined as
\begin{equation}
\begin{array}{c}
a^{S}_{c_ic_j}=
  \begin{cases}
    w^{S}_{c_ic_j},  & \quad \text{if } (c_i, c_j) \in E^S,\\
    0,  & \quad \text{otherwise}.\\
  \end{cases}\\
\end{array}
\end{equation}

\begin{figure}[t]
\centering
\includegraphics[width=0.75\linewidth]{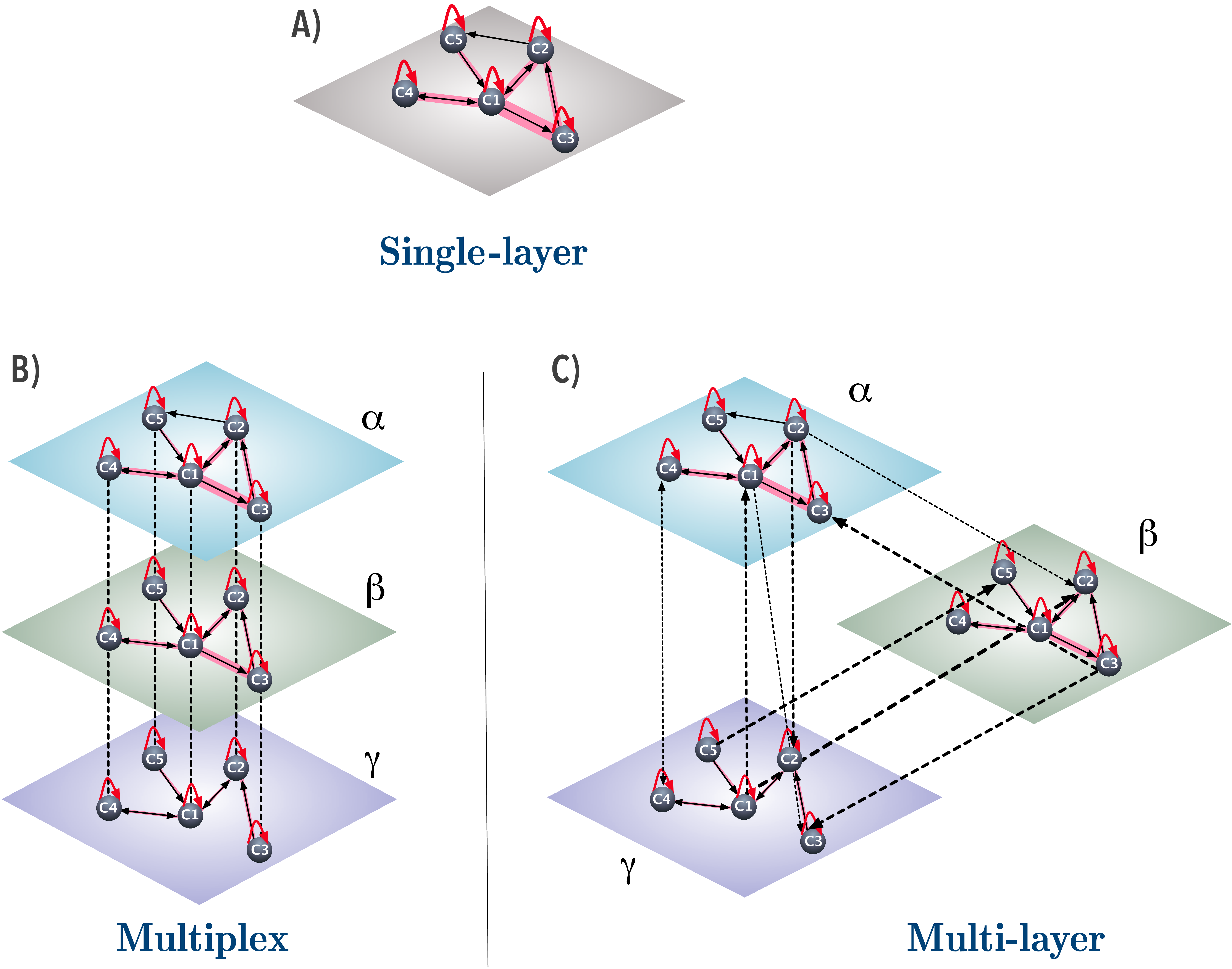}
\caption{Network representations of international trade. (\textbf{A}) the single-layer network of transactions among five countries. Each node $c_i$ represents a country, and the links (solid lines) refer to the interactions among countries. The weight $w_{c_ic_j}$ of each link (represented by the width of the line) from country $c_i$ to country $c_j$ is proportional to the sum of the value (US dollars) of all transactions of products and services exchanged from country $c_i$ to country $c_j$. Self-loops refer to economic transactions of a country with itself; (\textbf{B}) the multiplex network of transactions among countries. Each of the three layers is associated with an industry, and all layers are populated by the same nodes, each representing a country involved in transactions in the corresponding industry. In this representation, the directed connections within each layer (solid lines), i.e., the intra-layer connections, convey information regarding the amount of value exchanged from one country to another (or to itself) in the corresponding industry. Notice that the weights from the single-layer representation are distributed among layers in the multiplex network, and are represented by the different widths of the lines. Each cross-layer connection (dashed line) links each country to itself across layers and has no economic interpretation; (\textbf{C}) the multi-layer network of transactions among countries. This is a generalization of the two previous cases. Each layer represents an industry and all layers are populated by the same nodes, each representing a country. The intra-layer directed connections (solid lines) in each layer represent transactions from one country to another (or to itself) within the corresponding industry, while the cross-layer connections (dashed lines) represent transactions from one country to another (or to itself) across different industries. In the multi-layer network, the weights of the links from the multiplex network are re-distributed among connections both within and between layers, and are again represented by the different widths of the lines.}
\label{fig:networks}
\end{figure}

For example, let us consider a simplified version of the WIOT where there are only five countries ($c_1$, $c_2$, $c_3$, $c_4$, and $c_5$), and three different production sectors ($\alpha$, $\beta$, and $\gamma$). In the single-layer representation, all transactions from one country $c_i$ to another country $c_j$ across the different production sectors are collapsed into one single layer, and contribute to a single directed link $(c_i,c_j)$, with weight equal to the sum of all transactions from country $c_j$ to country $c_j$, as represented in Figure~\ref{fig:networks}A. Notice that, in this representation, we cannot distinguish between transactions that take place within (and~across) different production sectors. 

To properly distinguish between transactions according to the production stages from which they originate, we can introduce the concept of layer and represent the complex system of interactions among countries in different industries as a multiplex network~\cite{mastrandrea2014reconstructing}. In this network representation, $57$ layers, each representing a distinctive industry, are populated by $43$ countries. In each layer $\alpha$, a~connection is established from country $c_i$ to country $c_j$ if there is an economic transaction from country $c_i$ to country $c_j$ in layer $\alpha$. These within-layer transactions are referred to as intra-layer connections. Moreover, each country is assumed to be connected to itself across any two layers $\alpha$ and $\beta$, and these inter-layer connections represent the only couplings between industries. Formally, we can define the multiplex network as a pair $X=(G^{X},I^{X})$, where $G^{X}=\{G^{X}_{\alpha};~\alpha~\in~\{1,\dots,k\}\}$ is a family of directed graphs $G^{X}_{\alpha}=(V^{X}_{\alpha},E^{X}_{\alpha},W^{X}_{\alpha})$ associated with the layers of $X$, and $I^{X}$ is the set of interconnections between (the same) nodes in a different layers. Thus, the intra-layer adjacency matrix of each directed graph $G^{X}_\alpha$ will be denoted by $A^{X}_{[\alpha]}=a_{c_ic_j}^{X[\alpha]}$, where:
\begin{equation}
\begin{array}{c}
a_{c_ic_j}^{X[\alpha]} =
  \begin{cases}
    w_{c_ic_j}^{X[\alpha]},  & \quad \text{if } (c_i, c_j) \in E^{X}_\alpha,\\
    0,  & \quad \text{otherwise},\\
  \end{cases}\\
\end{array}
\end{equation}

\noindent for $1 \leq i, \; j \leq N$ and $1 \leq \alpha \leq k$, where $w_{c_ic_j}^{X[\alpha]}$ is the sum of the values of all transactions originating from country $c_i$ within a particular industry $\alpha$ and directed towards country $c_j$ in any other industry (including $\alpha$ itself).

In our simple example, the multiplex representation of trade implies that now there are three layers representing the different production sectors, and we can therefore distinguish between transactions among countries in different sectors (see Figure~\ref{fig:networks}B). Moreover, the weights of the transactions are now split among the sectors. For instance, in the single-layer case, the intensity $w_{c_1c_3}$ of the transaction from country $c_1$ to country $c_3$ is very strong and is represented by a very thick line (Figure~\ref{fig:networks}A), whereas, in the multiplex case, the same link connecting the two countries can be found in two distinct layers ($\alpha$~and $\beta$), and thus the intensity of the link in each layer varies. Notice that the sum of the weights in each layer is equal to the weight of the corresponding link in the single-layer representation, that is, $w^{S}_{c_ic_j}=w_{c_ic_j}^{X[\alpha]}+w_{c_ic_j}^{X[\beta]}+w_{c_ic_j}^{X[\gamma]}$. However, the cross-layer connections have no empirical interpretation as they simply serve the purpose of ensuring node alignment across layers. A multiplex network representation of international trade therefore neglects transactions that occur between different countries across different layers as well as transactions that involve the same country across layers.

The multi-layer network is indeed the only structure that can capture all such interactions among countries and industries and that can fully represent the information provided by the WIOT~\cite{alves2018nested}. As in the case of the multiplex network, in the multi-layer network, each layer represents a different industry, and each layer is populated by the same number of nodes representing the countries. The main difference between a multiplex and a multi-layer network lies in the way in which the connections are represented. In the multi-layer network, the intra-layer connections represent the transactions between countries that originate from, and terminate at, the same industry, whereas the cross-layer connections refer to the transactions that occur among countries across different sectors. Formally, we define a multi-layer network as a pair $M=(G^{M},C^{M})$, where $G^{M}=\{G^{M}_{\alpha};~\alpha~\in~\{1,\dots,k\}\}$ is a family of directed graphs $G^{M}_{\alpha}=(V^{M}_{\alpha},E^{M}_{\alpha},W^{M}_{\alpha})$ associated with the layers of $M$, and $C^{M}$ is the set of interconnections between nodes belonging to different graphs $G^{M}_{\alpha}$ and $G^{M}_{\beta}$ with $\alpha \neq \beta$. Formally, $C^{M}=\{C^{M}_{\alpha \beta};~\alpha,~\beta~\in~\{1,\dots,k\}, \alpha \neq \beta \}$ is a family of directed graphs $C^{M}_{\alpha \beta}=\{(c_i,c_j)\}$, where $\{i,j\, \in \{1,\dots, N\}\}$, $c_i \in V^{M}_{\alpha}$ and $c_j \in V^{M}_{\beta}$. Notice that, unlike the multiplex network, the multi-layer network allows the same country to be involved in economic transactions with itself across layers~(i.e.,~$i=j$). 

The definition of the multi-layer network is similar to the one of the multiplex network, except for the way in which the intra-layer weights and the cross-layer connections are defined. Formally, we can define the element $a_{c_ic_j}^{M[\alpha]}$ of the intra-layer adjacency matrix $A^{M[\alpha]}$ of each graph $G^{M}_\alpha$ as
\begin{equation}
\begin{array}{c}
a_{c_ic_j}^{M[\alpha]} =
  \begin{cases}
    w_{c_ic_j}^{M[\alpha]},  & \quad \text{if } (c_i,c_j) \in E^{M}_{\alpha},\\
    0,  & \quad \text{otherwise},\\
  \end{cases}\\
\end{array}
\end{equation}
where: $1 \leq i,j \leq N$; $1 \leq \alpha \leq k$; $c_i, c_j \in V^{M}_{\alpha}$; and $w_{c_ic_j}^{M[\alpha]}$ is the sum of the weights associated with all transactions originating from country $c_i$ within a particular industry $\alpha$ and directed to country $c_j$ within the same industry $\alpha$. {Thus, an intra-layer link between country $c_i$ and country $c_j$ in industry $\alpha$ is established when there is at least one transaction between $a_i$ and $a_j$ in $\alpha$.}

The element $a_{c_ic_j}^{M[\alpha \beta]}$ of the cross-layer adjacency matrix $A^{M[\alpha \beta]}$ corresponding to the set of interconnections $C^{M}_{\alpha \beta}$ can be defined as
\begin{equation}
a_{c_ic_j}^{M[\alpha \beta]} =
  \begin{cases}
     w_{c_ic_j}^{M[\alpha \beta]}, & \quad \text{if } (c_i, c_j) \in C^{M}_{\alpha \beta}, \\
    0,  & \quad \text{otherwise,}\\
  \end{cases}
\end{equation}
 where $1 \leq i, j \leq N$; $1 \leq \alpha, \beta \leq k$; $\alpha \neq \beta$; $c_i \in V^{M}_{\alpha}$; $c_j \in V^{M}_{\beta}$, and $w_{c_ic_j}^{M[\alpha \beta]}$ is the sum of the weights associated with all transactions originating from country $c_i$ within a particular industry $\alpha$ and directed towards country $c_i$ in industry $\beta$. {Thus, a cross-layer link between country $c_i$ in industry $\alpha$ and country $c_j$ in industry $\beta$ is established when there is at least one transaction between $a_i$ and $a_j$ across the corresponding industries.}

Figure~\ref{fig:networks}C illustrates the multi-layer network using our example of five countries and three industries. The weights are now re-distributed among the intra- and cross-layer connections and not only among the intra-layer connections as was the case with the multiplex network. This~enables us to extract more information from the WIOT, and properly capture the roles that countries play in the various production stages within the global value chain. Notice that, to recover the weights of the single-layer network, we need to sum the weights within and between layers---i.e.,~\mbox{$w^{S}_{c_ic_j}=\sum_\alpha \sum_\beta w_{c_ic_j}^{M[\alpha \beta]}$}. 

\section{Results}\label{section:results}

We started by computing the in- and out-strength distributions for each network. The in-strength of a node $c_i$ is defined as the sum of the weights $w_{c_jc_i}$ of all edges pointing to node $c_i$. Analogously, the out-strength of a node $i$ is defined as the sum of the weights $w_{c_ic_j}$ of all edges departing from node $c_i$. For the single-layer network, the in- and out-strengths of node $c_i$ can be formalized, respectively, as

\begin{equation}
 s_{c_i}^{S, in} = \sum_{c_j=c_1}^{c_N} w^{S}_{c_jc_i} \quad \text{and} \quad s_{c_i}^{S, out} = \sum_{c_j=c_1}^{c_N} w^{S}_{c_ic_j}. 
\end{equation}\label{eq:strength1}

Using the above definitions, we can compute the in- and out-strength sequences of the single-layer network of length equal to $43$, i.e., $L^{S,\,in}=L^{S,\,out}=43$. These sequences can then be used to obtain the strength distributions. To this end, we calculated the frequencies with which a given value of strength appears in the network by binning the data and computing the frequency of finding a specific value of strength in a given bin. 

\begin{figure}[t]
\centering
\includegraphics[width=0.9\linewidth]{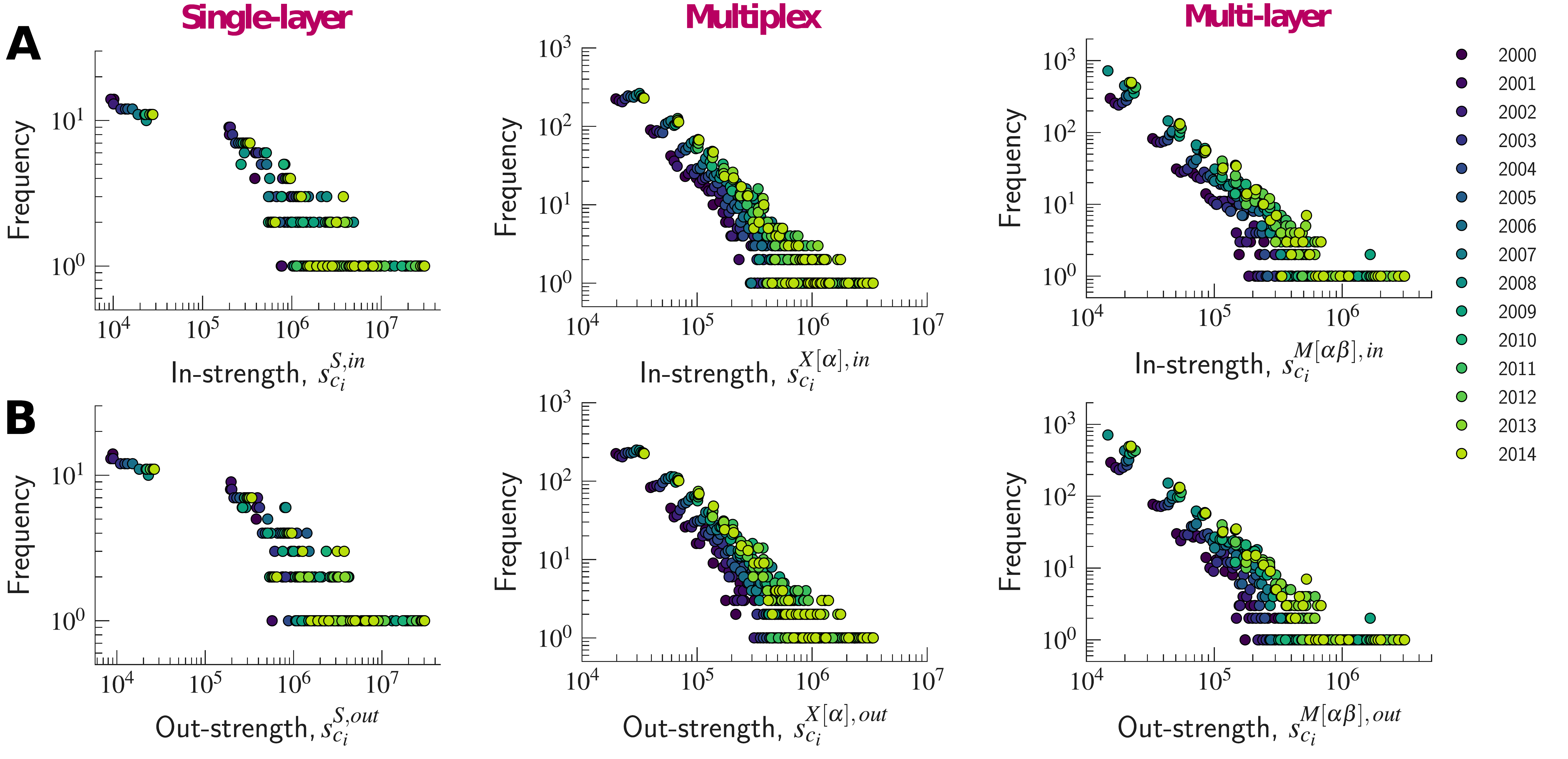}
\caption{Strength distributions of each network in the period from 2000 to 2014. The color varies from dark blue (year = 2000) to light green (year = 2014).  (\textbf{A}) from left to right, the upper panels show the in-strength distributions for the single-layer, multiplex, and multi-layer networks, whereas the bottom panels (\textbf{B}) show the out-degree distributions for the single-layer, multiplex, and multi-layer networks.}
\label{fig:distribution}
\end{figure}

Figure~\ref{fig:distribution} shows that, while there is no clear heavy-tailed behavior, the majority of nodes tend to have a small strength, whereas some have disproportionally large values. Moreover, this tendency towards heterogeneity seems to become more pronounced as time goes by. 

Next, we define node strength for the multiplex and multi-layer networks. In the multiplex network, each layer is associated with a directed graph $G^{X}_\alpha$ to which Equation~(\ref{eq:strength1}) can be applied to compute the strengths. Thus, for each layer $\alpha$, we can define the in- and out-strengths, $s_{c_i}^{X[\alpha],\,in}$ and $s_{c_i}^{X[\alpha],\,out}$ of node $c_i$, respectively, as

\begin{equation}
s_{c_i}^{X[\alpha],\,in}=\sum_{c_j=c_1}^{c_N}  w_{c_jc_i}^{X[\alpha]} \quad \text{and} \quad  s_{c_i}^{X[\alpha],\,out}=\sum_{c_j=c_1}^{c_N} w_{c_ic_j}^{X[\alpha]}.
\label{eq:strength2}
\end{equation}

By calculating $s_{c_i}^{X[\alpha],\,in}$ and $s_{c_i}^{X[\alpha],\,out}$ for every node $c_i$ in every layer $\alpha$, we obtain two strength sequences of maximum length $L^{X,\,in}$ and $L^{X,\,out}$, respectively, each equal to $43\times57$. Note that the actual length of the sequence may vary from year to year depending on whether some edges connecting countries within industries are absent from the network. Both sequences are therefore longer than the strength sequences obtained with the single-layer network \mbox{($L^{S,\,in}=L^{S,\,out}=43$)}.
In fact, in the multiplex network, we are able to measure the distinct contribution of each industry to the total strength of a country. One can also recover the strength sequences of the single-layer network by computing $s_{c_i}^{S,\,in}=\sum_\alpha s_{c_i}^{X[\alpha],\,in}$ and $s_{c_i}^{S,\,out}=\sum_\alpha s_{c_i}^{X[\alpha],\,out}$. In the case of the multiplex network, we can thus disentangle the strengths of countries across industries, and measure the contribution of each country to the total value exchanged in each industry. A simple plot of the distribution of strengths (Figure~\ref{fig:distribution}) shows that the node strengths concentrate within a much smaller range of values than was the case with the single-layer network. Nevertheless, the in- and out- strength distributions appear to be more heavy-tailed than the ones of the single-layer network, thus suggesting that countries' competitiveness and market shares differ from one industry to another.

While the multiplex network enables us to uncover the contributions of each country to the value exchanged in each industry, it does not offer information on each country's contribution to the value exchanged between industries. To rectify this shortcoming, we constructed the multi-layer network in which edges can connect countries both within and across layers. 
To properly capture the distinction between these two types of edges, the in- and out-strengths, $s_{c_i}^{M[\alpha \beta],\,in}$ and $s_{c_i}^{M[\alpha \beta],\,out}$, of node $c_i$ can be formally defined, respectively, as
\begin{equation}
 s_{c_i}^{M[\alpha \beta],\,in}=\sum_{c_j=c_1}^{c_N}  w_{c_jc_i}^{M[\beta \alpha]} \quad \text{and}  \quad  
 s_{c_i}^{M[\alpha \beta],\,out}=\sum_{c_j=c_1}^{c_N}  w_{c_ic_j}^{M[\alpha \beta]}.
\label{eq:strength3}
\end{equation}

By calculating $s_{c_i}^{M[\alpha, \beta],\,in}$ and $s_{c_i}^{M[\alpha \beta],\,out}$ for every node $c_i$ and for every pair of layers $\alpha \beta$, we obtain two strength sequences of maximum length $L^{M,\,in}$ and $L^{M,\,out}$, respectively, each equal to $43\times57\times57$. Once again, the actual length of the sequence may vary from year to year because some edges connecting some countries in some industries to other countries in other industries might be absent from the network. Both sequences are therefore longer than the strength sequences obtained with the multiplex network ($L^{M,\,in}=L^{M,\,out}=43\times57$) and the single-layer network ($L^{S,\,in}=L^{S,\,out}=43$).

Given the above definitions, we can easily recover the node strength sequences of the multiplex network by summing the strengths over the layers, i.e., $s_{c_i}^{X[\alpha],\,in} = \sum_\beta s_{c_i}^{M[\alpha \beta],\,in}$ and $s_{c_i}^{X[\alpha],\,out} = \sum_\beta s_{c_i}^{M[\alpha \beta],\,out}$. Moreover, we can recover the strength sequences of the single-layer network by summing the strengths over all pairs of layers, i.e., $s_{c_i}^{S,\,in} = \sum_\alpha \sum_\beta s_{c_i}^{M[\alpha \beta],\,in}$ and $ s_{c_i}^{S,\,out} = \sum_\alpha \sum_\beta s_{c_i}^{M[\alpha \beta],\,out}$. 

Using the definition given by Equation~(\ref{eq:strength3}), we can distinguish between strengths associated with different layers as well as between strengths originated from transactions occurring within the same layer and  transactions between different layers along the global value chain. 
Uncovering distinct weights for different types of edges has the effect of changing the way the weights are distributed among edges and, as a result, might affect the shape of their distributions. Indeed, Figure~\ref{fig:distribution} suggests that the multi-layer network is characterized by values of node strength that (by construction) concentrate within a much smaller range than was the case with the single-layer and the multiplex networks. However, most interestingly, in the case of the multi-layer network, the strength distributions seem to exhibit a clearer heavy-tailed behavior than was the case with the distributions in the other two networks. 

A more suitable topology for representing the trade system, such as the multiplex or the multi-layer network, can provide us with a more detailed description of the interactions among countries and industries. In particular, a multi-layer representation can offer new insights into the role each country plays at the local level, in each industry, and at the same time it can unveil properties of the network at the global level, as suggested by the strength distribution with different ranges of values. Thus, by changing the level of analysis of the interactions among countries and industries, from the single-layer to the multiplex and the multi-layer networks, we are able to extract and analyze more information about production stages, the structure of the global value chain, and the different roles of countries and products in the international production network.

To quantify the information gained from one representation to another, in what follows we shall propose to evaluate a metric for characterizing and comparing the diversity of each network. Traditionally, a measure that has been widely used to describe the distribution of a given amount of a resource across the various elements of a system is the Shannon entropy of the resource~\cite{battiston2014structural}. In~our case, the resource is the total value exchanged among the countries, and the elements across which this value can be distributed are the various (combinations of) transactions among countries. Thus, entropy aims to measure the extent to which the total traded value is uniformly distributed among transactions or rather it is concentrated on a small set of economic transactions. Moreover, since the level of granularity varies across the various network representations of trade, we would expect entropy to take on different values reflecting the different distributions of values across countries and industries. Formally, the entropies $H^{in}$ and $H^{out}$ of the in- and out-strengths observed in a trade network can be defined, respectively, as
\begin{equation}
H^{in}=-\sum_{l=1}^{L^{in}} p_l^{in} \log{p_l^{in}} \quad \text{and}  \quad 
H^{out}=-\sum_{l=1}^{L^{out}} p_l^{out} \log{p_l^{out}}, 
\end{equation}\label{eq:entropy}
where 
\begin{equation}
p_l^{in}=\frac{s_l^{in}}{\sum_{h=1}^{L^{in}} s_l^{in}} \quad \text{and} \quad
p_l^{out}=\frac{s_l^{out}}{\sum_{h=1}^{L^{out}} s_l^{out}}.
\end{equation}

 In particular, we have: (i) $s_l^{in}=S_{c_i}^{S,\,in}$, $s_l^{out}=S_{c_i}^{S,\,out}$, and $L^{in}=L^{out}=L^{S,\,in}=L^{S,\,out}=43$ in the case of the single-layer network; (ii) $s_l^{in}=S_{c_i}^{X[\alpha],\,in}$, $s_l^{out}=S_{c_i}^{X[\alpha],\,out}$, and $L^{in}=L^{out}=L^{X,\,in}=L^{X,\,out}=43\times57$ in the case of the multiplex network; and (iii) $s_l^{in}=S_{c_i}^{M[\alpha \beta],\,in}$, $s_l^{out}=S_{c_i}^{M[\alpha \beta],\,out}$, and
$L^{in}=L^{out}=L^{M,\,in}=L^{M,\,out}=43\times57\times57$ in the case of the multi-layer network.

So conceived of, the entropy of in- or out-strength captures the degree to which in- or out-strength is uniformly distributed across the system~\cite{aleta2018multilayer,battiston2014structural}. In particular, entropy is equal to zero in the limiting case where the entire value traded in the system is concentrated on one strength. For example, this is the (unrealistic) case in which only one country imports or exports the whole value from or to other countries. Conversely, entropy takes its maximum value ($log{L^{in}}$ or $log{L^{out}}$) when the entire value is uniformly distributed over the different strengths. Therefore, the higher the value of entropy, the more uniformly the total value traded in the system is distributed across the various in- and out-strengths. {Comparing the entropies of distributions of different event spaces characterized by a different number of possible outcomes could be problematic, precisely because entropy scales with such number of outcomes and can thus take on different maximum values. Therefore, because our aim was to compare entropies across the three different networks characterized by three different event spaces, we normalized each entropy} by dividing it by its possible maximum value $log{L^{in}}$ or $log{L^{out}}$ in the corresponding network.

\begin{figure}[t]
\centering
\includegraphics[width=0.9\linewidth]{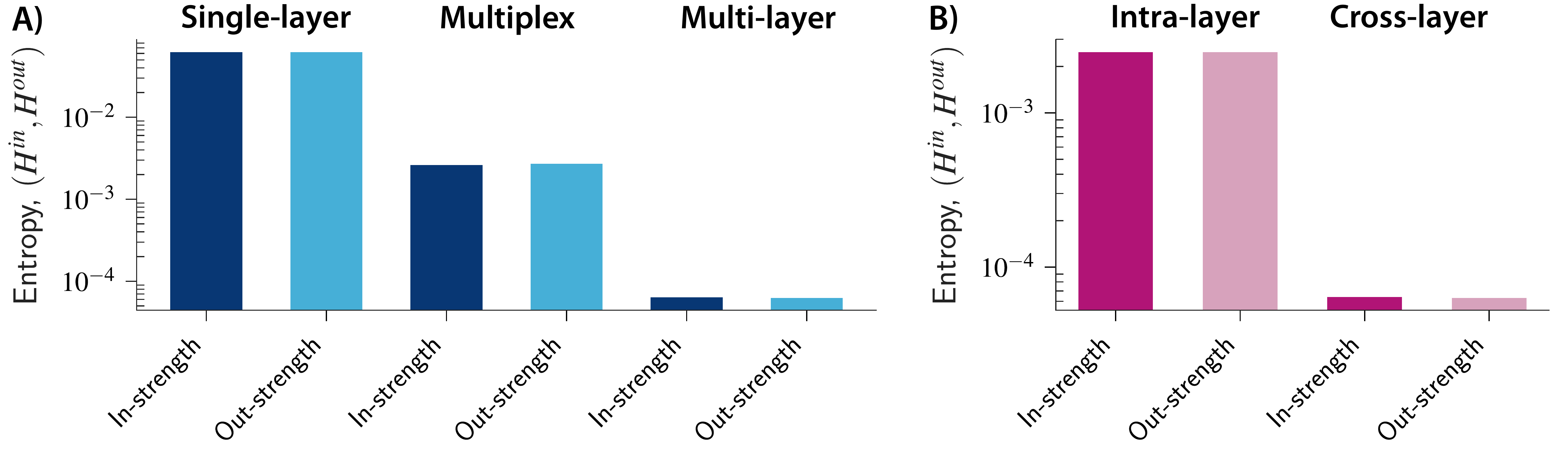}
\caption{Entropy of node strength in the worldwide single-layer, multiplex, and multi-layer networks in 2000. (\textbf{A}) As we increase the complexity of the trade structure from the single-layer to the multi-layer network, strength is re-distributed less uniformly across transactions and stages of production, and new heterogeneous market structures and dominance positions are uncovered; (\textbf{B}) entropy of the in- and out- strengths associated distinctively with intra- and cross-layer connections in the multi-layer network. Cross-layer connections are characterized by a lower value of entropy, thus suggesting a higher concentration of value in production stages that span different industries than in production stages that occur within industries. 
}\label{fig:entropy}
\end{figure}

Figure~\ref{fig:entropy}A shows the entropy for the in- and out-strengths in the three networks in 2000. As~we move from a simple representation to a more complex structure, the entropy decreases, thus suggesting that a small set of economic transactions tend to attract a disproportionally large amount of the total value traded in the system. 
At the same time, transactions become more similar to each other and the total value becomes more uniformly distributed at a global level when transactions are aggregated within industries (in the multiplex network) and within countries (in the single-layer network). 
Moreover, the multi-layer representation allows us to distinguish between strengths related to intra-layer connections and strengths associated with cross-layer transactions between different industries. Figure~\ref{fig:entropy}B shows the entropies of the in- and out-strengths related to intra- and cross-layer connections in 2000. The results suggest that there is much more diversity in the way value is distributed across the cross-layer connections (transactions among different industries) than across the intra-layer connections (transactions within industries). Thus, by further increasing the granularity of the level of analysis, we can uncover heterogeneous market structures that would otherwise remain concealed at a more aggregate level. 

\begin{figure}[t]
\centering
\includegraphics[width=0.9\linewidth]{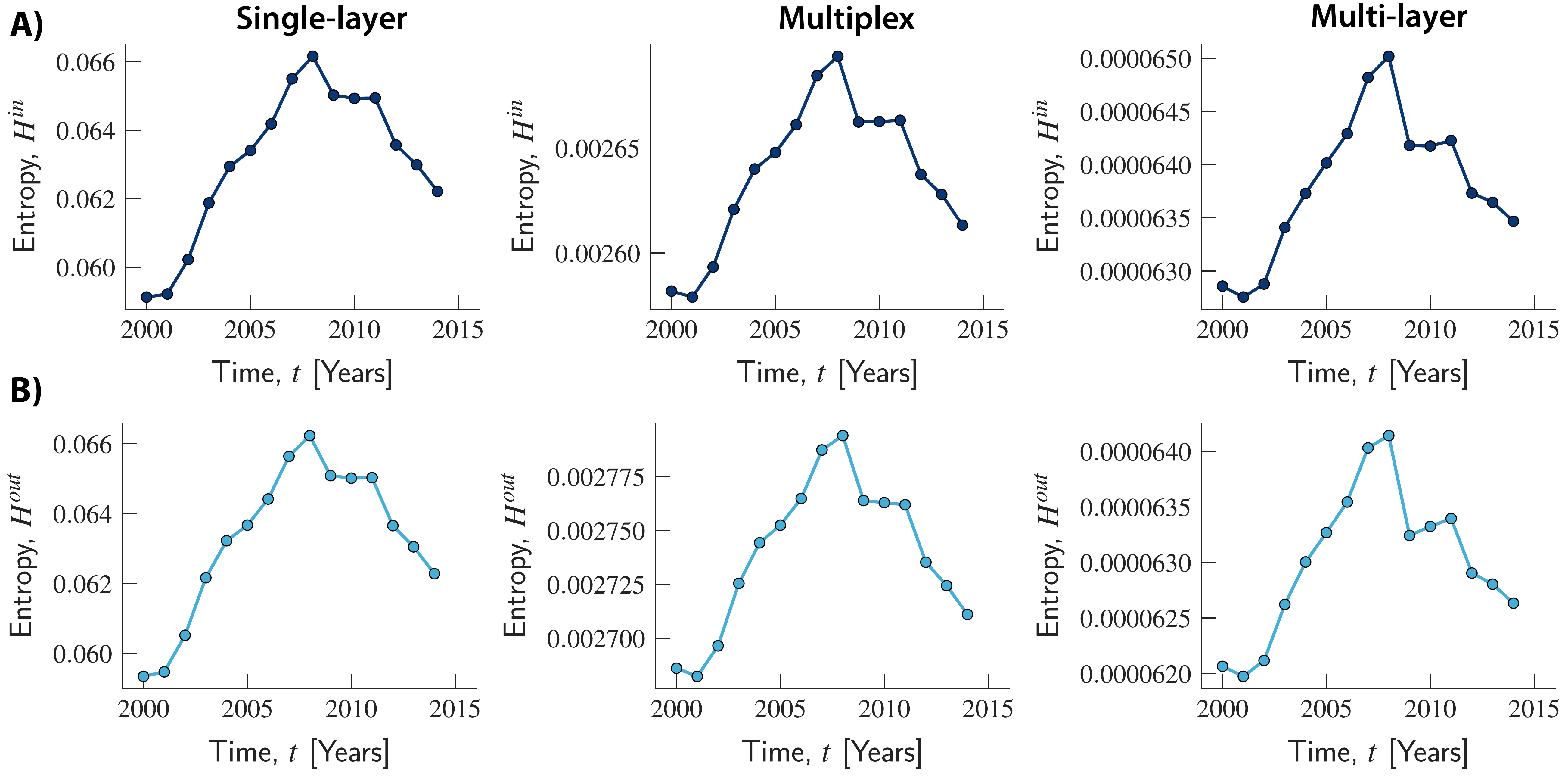}
\caption{Evolution of entropy of strength in the international trade network in each network representation. (\textbf{A}) the upper panels show the evolution of the entropy of in-strength in the single-layer (left-hand column), multiplex (middle column), and multi-layer (right-hand column) networks; (\textbf{B})~the bottom panels show the evolution of the entropy of out-degree strength in the three respective~networks. }\label{fig:entropyevolution}
\end{figure}

Next, we shall focus on how entropy changed over time, and whether different evolution trends can be uncovered in the three network structures. To this end, we calculated the entropy of the in- and out-strengths for the single-layer, multiplex and multi-layer networks for every year from 2000 to 2014. Figure~\ref{fig:entropyevolution} suggests that a consistent trend can be observed across the three networks. In 2001, entropy began to increase steadily over time, reaching its local maximum in 2008, at a time in which the global economic downturn was at its peak. In 2009, entropy started a declining trend that continued over time for the whole observation period (apart from a slight increase in 2010 and 2011 in the multiplex and multi-layer networks). Moreover, across all years, the relationship among the entropies of the three networks is consistent with the findings in Figure~\ref{fig:entropy}. Over time, entropy in the multi-layer network remained at lower values than in the multiplex network, which in turn was characterized by lower values than the single-layer network.

\begin{figure}[t]
\centering
\includegraphics[width=0.8\linewidth]{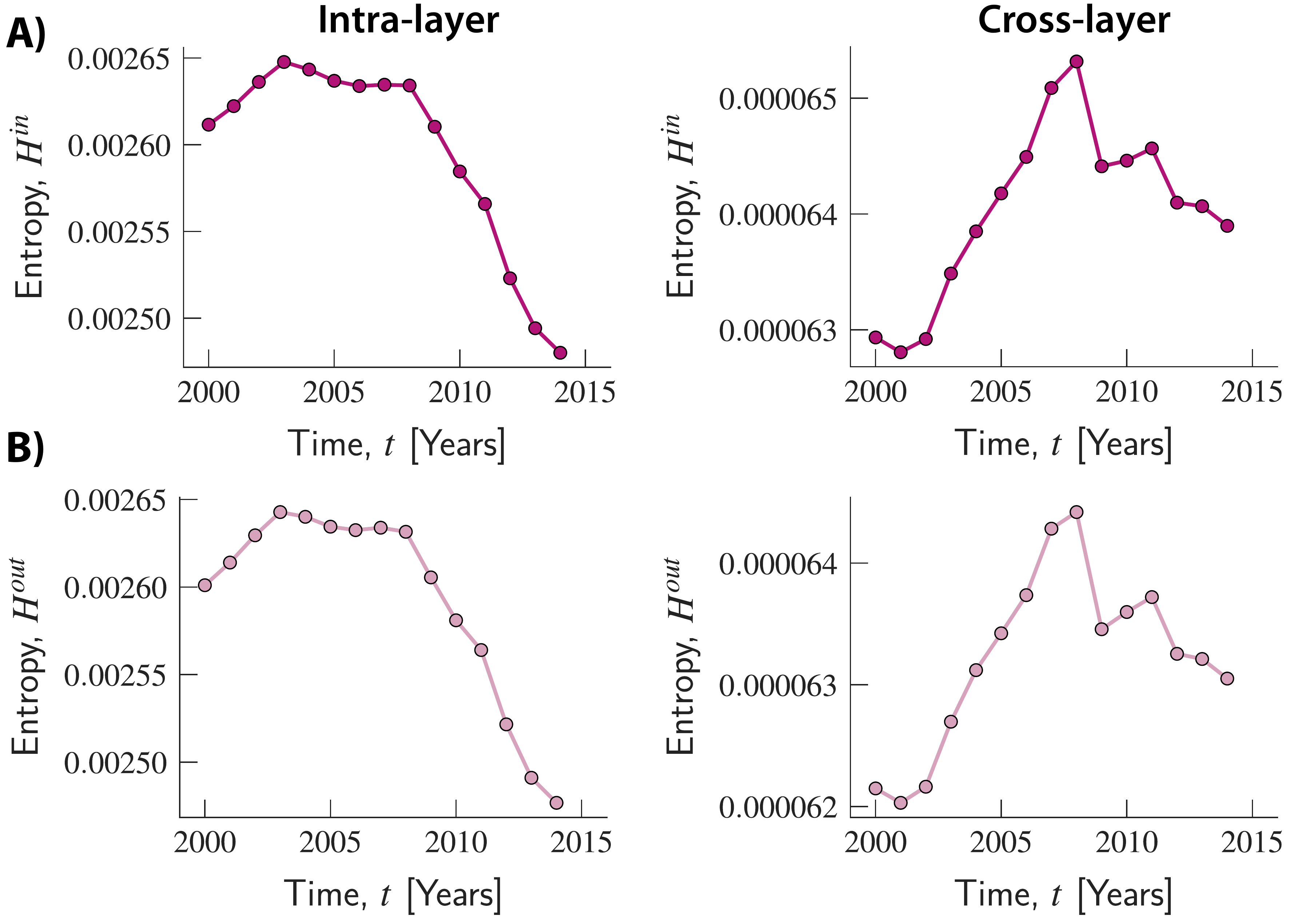}
\caption{Evolution of the entropy of in- and out-strengths related to the intra- and cross-layer connections in the multi-layer network. (\textbf{A}) The upper panels show the evolution of entropy of in-strength related to intra-layer (left-hand column) and cross-layer (right-hand column) connections. (\textbf{B}) The bottom panels show the evolution of entropy of out-strength associated with the two types of connections.}\label{fig:entropyevolutionlayers}
\end{figure}

To further explore the evolution of entropy in the multi-layer representation, and the interplay between its components, we computed the entropies related to the intra- and cross-layer strengths. Figure~\ref{fig:entropyevolutionlayers} suggests that what is responsible for the increasing trend of the overall entropy observed in the multi-layer network is the component associated with cross-layer transactions. Indeed, while this increased over time until 2008, the component related to intra-layer connections slightly increased up to 2003 and then remained fairly stable until 2008. Both components then started a declining trend, although the decline was much more pronounced for the entropy related to intra-layer connections which therefore seems to be the driving force underlying the decline in global entropy. Moreover, across all years, the entropy related to intra-layer connections took values that are comparable to those observed in the multiplex network. By contrast, values of entropy related to cross-layer connections were always at a lower scale, and can therefore be seen as responsible for the lower total values compared to the ones observed in the other two networks. Once again, using a finer-grained network representation of trade can unmask market structures and sources of heterogeneity in the system that would otherwise remain undetected if values were aggregated across transactions and simpler (and less accurate) structures were used.

\section{Discussion}\label{section:dissussion}

International trade among countries can be investigated from different perspectives and at different levels of detail~\cite{allard2017geometric,caldarelli2012network,caldarelli2018physics,cingolani2017countries,cristelli2015heterogeneous,
de2011world,fagiolo2009world,formichini2018measuring,garlaschelli2004fitness,garlaschelli2005structure,he2010structure,
hidalgo2007product,hidalgo2009building,serrano2003topology,tacchella2012new}. Traditionally, a network-based perspective has been advocated to represent countries as nodes and economic transactions as directed edges connecting nodes~\cite{de2011world}. A~network representation of trade has often resulted from one-mode projections of bipartite networks in which a set of nodes represents the countries and another set includes
the industries to (from) which the countries export (import)~\cite{cristelli2013measuring,saracco2015randomizing}. However, neither a bipartite network connecting countries to
(exported or imported) products nor its one-mode projections, 
such as the product-to-product or country-to-country networks, can account for the full extent of interactions that typically occur among countries in the international production network underpinning the global value chain. This article was aimed precisely at assessing how, by gradually increasing the complexity of the network framework, novel insights can be gained on the structure of interactions among countries as well as on the underlying forces driving the evolution of the global value chain. 

{Our study was based on the WIOD, which provides the most complete publicly available representation of trade between countries and industries. This enabled us to investigate the structure of connections between countries in the international production network. Notice that, while this data set comes with the price of restricting the analysis to a limited sample of countries and industries, all other available data sets that account for more countries and industries (e.g., the United Nations COMTRADE data) do not provide as detailed information as the WIOD on single economic transactions and cannot therefore be used for the analysis of the global value chain.}  

We focused on three network frameworks---the single-layer, multiplex, and multi-layer networks---and investigated the extent to which the total traded value is distributed within these three networks. To this end, we computed the in- and out-strengths of countries, defined as the total value imported or exported by the countries and associated with economic transactions that can be aggregated in various ways by origin or destination. In the single-layer network, transactions are aggregated by country. In this case, a country's in- or out-strength reflects the whole value of transactions, respectively, originated from or directed to all other countries (including itself). In the multiplex network, transactions are aggregated by industry from which they originated and by country. In this case, a country's in- or out-strength in a given industry is defined as the sum of values of all transactions departing from that industry and, respectively, originating from or directed to all other countries (including the focal country itself). Finally, in the multi-layer network, a finer-grained structure results from aggregating transactions by country and combinations of industries. In this case, a country's in- or out-strength in a given pair of industries is defined as the sum of all values of transactions occurring between the two industries and, respectively, originating from or directed to all other countries. Thus, as we move from a single-layer network representation to more complex ones, the weights of links are re-distributed over different layers in the multiplex representation and over layers and combinations of layers in the case of the multi-layer network. 

Findings suggest that the re-allocation of weights resulting from the use of more complex network representations causes the distribution of strengths to vary, making the heavy-tailed shape of the distribution appear more clearly in the case of more complex topologies. Thus, what may appear as a uniform market structure at an aggregate level can instead be represented as a more heterogeneous system where the total value traded is re-distributed among transactions by industry or by production stage. We also observed that the entropy of countries' strengths systematically decreases from the single-layer representation to the multiplex and multi-layer ones across all years. While countries may seem to play similar roles when trade is aggregated across industries, heterogeneity can be uncovered when a more suitable network framework enables us to assess the countries' contributions to the various production stages along which intermediary products are transformed into final ones. 

In addition, we studied the evolution of entropy over time, and detected an increasing trend of entropy up to a maximum in 2008, when the global economic downturn reached its peak. {These results contribute to the literature and current debates on structural changes in the economic system that can be regarded as the topological precursors and early-warning signals of an approaching crisis. For example, recent work has focused on a family of bipartite motifs and uncovered two phases in the evolution of the topological randomness of international trade. It has been suggested that the system began to exhibit increasing degrees of randomness and to weaken its internal structure four years before the onset of the financial crisis in 2007, and then reached a stationary regime characterized by a constant level of randomness during the following three years~\cite{saracco2016detecting}. Similarly, a recent study of the Dutch interbank network from 1998 to 2008 has suggested that, while the size and density of the network are uninformative about the crisis, a number of higher-order topological properties of the network (e.g., dyadic and triadic motifs) underwent a slow and continuous transition that started three years in advance of the crisis~\cite{squartini2013early}. In qualitative agreement with these studies, our work has also uncovered a pre-crisis build-up phase characterized by structural changes of the network related to the way the total traded value was distributed among transactions between countries and across industries (i.e., changes in the entropy of in- and out-strength). These changes in entropy can indeed be seen as warning signals of the upcoming economic downturn.}

{We further investigated these structural changes in the run-up to the crisis, by focusing on the multi-layer representation and by separating the contributions to strength} that resulted from the different types of transactions, namely those occurring within industries and those between industries. Interestingly, the entropy related to cross-layer connections remained smaller over the years than the entropy associated with intra-layer connections, and while the former was responsible for the overall increase in global entropy up to 2008, the latter was the component that most contributed to the subsequent overall decline. Distinguishing between transactions within the global value chain can therefore shed a new light on the sources of heterogeneity in the system. 

\section{Conclusions}

In summary, our study proposed a general framework for capturing and comparing the intricacies of three increasingly complex network structures. By focusing on international trade and on two specific measures---strength and entropy---we showed how fundamental nuances of the underlying structure may remain undetected when values and interactions are aggregated and a simplified yet unrealistic representation of the system is used. Because a variety of other real-world systems, beyond trade, can also be represented in terms of the single-layer, multiplex, and multi-layer network topologies, our work can inspire a number of empirical applications and has implications for other fields within and beyond the natural and social sciences. 

\section{Author contributions}
Conceptualization, L.G.A.A., G.M., F.A.R., P.P., and Y.M.; Methodology, L.G.A.A., G.M., P.P., and Y.M.; Software, L.G.A.A., G.M., F.A.R., P.P., and Y.M.; Validation, L.G.A.A., G.M., F.A.R., P.P., and Y.M.; Formal Analysis, L.G.A.A., G.M., F.A.R., P.P., and Y.M.; Investigation, L.G.A.A., G.M., F.A.R., P.P., and Y.M.; Resources, L.G.A.A., G.M., F.A.R., P.P., and Y.M.; Data Curation, L.G.A.A., G.M., F.A.R., P.P., and Y.M.; Writing---Original Draft Preparation, L.G.A.A., G.M., F.A.R., P.P., and Y.M.; Writing---Review and Editing, L.G.A.A., G.M., F.A.R., P.P., and Y.M.; Visualization, L.G.A.A., G.M., P.P., and Y.M.; Supervision, L.G.A.A., G.M., F.A.R., P.P., and Y.M.; Project Administration, L.G.A.A., G.M., F.A.R., P.P., and Y.M.

\section{Funding}
L.G.A.A. acknowledges S\~ao Paulo Research Foundation (FAPESP)(Grant No. 2016/16987-7) for financial support. F.A.R. acknowledges  National Council for Scientific and Technological Development (CNPq) (Grant No. 305940/2010-4) and S\~ao Paulo Research Foundation (FAPESP) (Grants No. 2016/25682-5 and Grants 2013/07375-0). Y.M. acknowledges support from the Government of Arag\'on, Spain through Grant E36-17R, by Ministerio de Econom\'ia y Competitividad (MINECO) and Fondo Europeo de Desarrollo Regional (FEDER) funds (Grant FIS2017-87519-P) and by the European Commission FET-Proactive Project Dolfins (Grant 640772).

\section{Conflicts of interest}
The authors declare no conflict of interest.

\end{document}